\documentclass[%
 reprint,
 amsmath,amssymb,
 aps,
prb,
]{revtex4-2}

\usepackage{graphicx}
\usepackage{dcolumn}
\usepackage{bm}

\begin{document}

\preprint{APS/123-QED}

\title{Systematic study of multi-magnon binding energies in the FM-AFM $J_1$-$J_2$ chain}

\author{Satoshi Nishimoto}
\affiliation{Department of Physics, Technical University Dresden, 01069 Dresden, Germany}
\affiliation{Institute for Theoretical Solid State Physics, IFW Dresden, 01069 Dresden, Germany}

\date{\today}

\begin{abstract}
We present a systematic study of multi-magnon bound states (MBSs) in the spin-$\tfrac{1}{2}$ FM-AFM $J_1$-$J_2$ chain under magnetic fields using the density-matrix renormalization group method. As a quantitative measure of stability, we compute the magnon binding energy $E_{\rm b}(M,p)$ for bound clusters of size $p$ over wide ranges of the frustration ratio $J_2/|J_1|$ and the normalized magnetization $M/M_{\rm s}$. Near saturation, we benchmark our data against the analytic two-magnon result and map out a clear hierarchy of $p$-magnon states, whose phase boundaries follow an empirical scaling $J_{2,{\rm c}}(p;p\!+\!1)/|J_1|\!\approx\!0.34\,p^{-2.3}$ for large $p$. We further quantify the relation between the most stable $p$ and the zero-field pitch angle $\theta$, verifying the conjectured inequality $1/p>\theta/\pi>1/(p+1)$ up to $p \lesssim 9$. The binding energy shows pronounced suppression as $J_2/|J_1|\!\to\!1/4^+$ and, for some frustration values, attains a maximum below full saturation, indicating that partial depolarization enhances bound-magnon mobility. Close to the FM instability, $E_{\rm b}(M_{\rm s},p)$ exhibits an empirical power-law vanishing consistent with a quantum-Lifshitz scenario. Our results provide a comprehensive, experimentally relevant map of MBS stability across field and frustration, offering concrete guidance for inelastic probes in quasi-one-dimensional magnets.
\end{abstract}

\maketitle


\section{\label{sec:intro}Introduction}

Multi-magnon bound states (MBSs) were predicted nearly a century ago~\cite{Bethe1931} and have since remained a central theme in quantum magnetism due to their striking many-body properties~\cite{Penc2011,Starykh2015}. In recent years, the emergence of MBSs in frustrated ferromagnetic-antiferromagnetic (FM-AFM) $J_1$-$J_2$ spin chains, which are widely used as minimal models for quasi-one-dimensional (Q1D) edge-shared cuprates~\cite{Kuzian2023}, has been extensively investigated on the theoretical side~\cite{Chubukov1991,Heidrich-Meisner2006,Vekua2007,Kecke2007,Hikihara2008,Sudan2009,Zhitomirsky2010,Sato2013,Fukuhara2013,Onishi2015,Schaefer2020}. In parallel, considerable experimental efforts have sought signatures of such bound states in highly 1D compounds, including  LiCuVO$_4$~\cite{Svistov2011,Mourigal2012,Nawa2013,Buettgen2014,Orlova2017}, LiCuSbO$_4$~\cite{Dutton2012,Grafe2017,Bosiocic2017}, and linarite  PbCuSO$_4$(OH)$_2$~\cite{Effenberger1987,Yasui2011,Willenberg2012,Thermodynamic2013,Willenberg2016}, where features consistent with spin-nematic or more general multipolar correlations have been reported. Nevertheless, an unambiguous realization in nature remains challenging, particularly in Q1D quantum magnets, because stabilizing MBSs has traditionally required high magnetic fields; in addition, they can be fragile against thermal fluctuations and external perturbations~\cite{Ueda2009,Interplay2015}.

At zero magnetic field, the spin-$\tfrac{1}{2}$ FM-AFM $J_1$-$J_2$ chain exhibits a fully polarized FM ground state for weak frustration ($J_2/|J_1|<1/4$). For convenience, we define the frustration ratio as $\alpha = J_2/|J_1|$ hereafter. At the critical point $\alpha = 1/4$, the FM state becomes unstable and gives way to nontrivial singlet ground states~\cite{Bader1979,Haertel2008}. Beyond this point, field-theoretical and numerical studies predicted spiral (incommensurate) correlations with dimerization tendencies~\cite{Bursill1995,Sirker2011,Furukawa2012} and, at large frustration, an exponentially small spin gap~\cite{Nersesyan1998,Itoi2001}. Subsequent high-accuracy calculations established that a finite but extremely small gap indeed opens for $\alpha>1/4$~\cite{Agrapidis2019}, peaking at only $\Delta \sim 0.007|J_1|$ near $\alpha\simeq0.6$. The ground state in this regime is best described as a third-neighbor valence-bond solid (D$_3$–VBS) selected by the order-by-disorder mechanism~\cite{Agrapidis2019}, with clear signatures of symmetry-protected topological order~\cite{Pollmann2010,den_Nijs1989} akin to the spin-1 AFM Heisenberg chain~\cite{Haldane1983-1,Haldane1983-2} and the Affleck–Kennedy–Lieb–Tasaki chain~\cite{Affleck1987}. This interpretation is supported by the presence of a finite string order parameter, a characteristic fourfold entanglement-spectrum degeneracy, and long-range third-neighbor dimer correlations~\cite{Agrapidis2019}. Thus, the FM–AFM $J_1$–$J_2$ chain provides a paradigmatic example in which frustration and quantum fluctuations cooperate to stabilize a gapped topological phase at zero field.

Under magnetic fields, the system enters magnetized states where flipped spins (magnons) may interact attractively. These effective interactions can stabilize MBSs not only in the dilute limit near saturation but also at intermediate magnetizations, giving rise to a variety of multipolar phases~\cite{Hikihara2008,Sudan2009}. When a field is applied to the spiral regime, a vector-chiral phase appears, characterized by spontaneous breaking of discrete parity ($Z_2$) with a finite longitudinal vector chirality, $\kappa_l^{(n)}=\langle (\vec S_l \times \vec S_{l+n})^z\rangle$~\cite{Hikihara2008}. With increasing field, $p$-magnon bound states emerge whose size $p$ depends sensitively on $\alpha$, leaving characteristic fingerprints in magnetization processes and excitation spectra~\cite{Kecke2007,Sudan2009}.

A natural quantitative measure of the stability of such bound states is the {\it magnon binding energy}, which can appear experimentally as the lowest transverse excitation gap in inelastic neutron scattering spectra~\cite{Onishi2015,Clio2025}. From the viewpoint of prospective materials and measurements, it is therefore important to know how robust MBSs are as a function of field-induced magnetization. Recently, it has even been proposed that internal fields may induce finite magnetization and thereby realize MBSs without applying an external field~\cite{Clio2025}; however, insofar as binding energies are primarily governed by magnetization (and frustration), their stability criteria should be essentially analogous in both situations. Despite several case studies, to our knowledge there has been no comprehensive mapping of the binding energy over broad ranges of frustration and magnetization parameters~\cite{J_Phys_Conf_Ser_400_032069,Onishi2015,Interplay2015}.

In this work, we provide a systematic determination of the magnon binding energy $E_{\rm b}(M,p)$ -- the energy gain of a $p$-magnon bound cluster relative to unbound magnons -- in the spin-$\tfrac{1}{2}$ FM-AFM $J_1$-$J_2$ chain using the density-matrix renormalization group (DMRG) calculations. We chart $E_{\rm b}(M,p)$ across wide ranges of the frustration ratio $\alpha$ and the normalized magnetization $M/M_{\rm s}$ (where $M_{\rm s}$ is the saturation magnetization) for bound clusters containing $p$ magnons. We further elucidate the near-saturation regime by quantifying the relationship between the most stable cluster size $p$ and the zero-field pitch angle, and by identifying the hierarchy and stability windows of multipolar (nematic/triatic/quartic, etc.) states. Our results provide concrete, experimentally relevant benchmarks for the observation of MBSs in Q1D magnets.

\section{\label{sec:model}Model and method}

\begin{figure}[t]
	\includegraphics[width=0.8\linewidth]{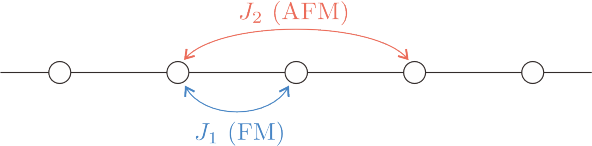}
	\caption{\label{fig:lattice}Lattice structure of the FM-AFM $J_1$-$J_2$ chain, where each circle denotes a spin-$\tfrac{1}{2}$ site.}
\end{figure}

In this section, we present the model Hamiltonian and the theoretical framework used to analyze the formation and stability of MBSs in the FM-AFM $J_1$-$J_2$ chain. Section~II.A introduces the spin-$\tfrac{1}{2}$ Hamiltonian in the presence of an external magnetic field. In Sec.~II.B we briefly describe the notion of magnon binding, which provides the basis for understanding the emergence of bound states at finite magnetization. Sec.~II.C defines the magnon binding energy, the central quantity of our analysis. Finally, Sec.~II.D outlines the details of our DMRG calculations.

\subsection{\label{subsec:hum}Hamiltonian}

The Hamiltonian of the spin-$\tfrac{1}{2}$ FM–AFM $J_1$-$J_2$ chain in the presence of an external magnetic field is given by
\begin{align}
	{\cal H}_\alpha=J_1\sum_i\vec{S}_i \cdot \vec{S}_{i+1}+J_2\sum_i\vec{S}_i \cdot\vec{S}_{i+2}-h\sum_i S^z_i,
	\label{eq:1Dham}
\end{align}
where $\vec{S}_i=(S^x_i,S^y_i,S^z_i)$ denotes the spin-$\tfrac{1}{2}$ operator (equivalent to the Pauli matrices) at site $i$, and $S^\pm_i=S^x_i \pm iS^y_i$. 
Here, $J_1\!<\!0$ and $J_2\!>\!0$ represent the FM nearest-neighbor and AFM next-nearest-neighbor exchange interactions, respectively, while $h$ denotes the external magnetic field. This frustrated spin chain is often used as a minimal model for Q1D edge-shared cuprates, where the nearest-neighbor interaction is typically FM due to the direct overlap of Cu–O orbitals along the edge~\cite{Kuzian2023}. We use $|J_1|=1$ as the energy unit.

\subsection{\label{subsec:magnon}Magnon binding}

Magnons are quantized spin-wave excitations corresponding to local spin flips on top of a magnetically ordered, polarized, or short-range correlated background. In 1D frustrated systems, magnons interact strongly and their mutual interactions can become effectively attractive. Consequently, several magnons may bind together to form composite excitations. The simplest example is a two-magnon bound state (bimagnon), which is directly related to spin-nematic (quadrupolar) correlations and corresponds to $p=2$. Under appropriate conditions, larger bound states involving three or more magnons can also be stabilized, leading to multipolar phases—triatic ($p=3$), quartic ($p=4$), etc.—characterized by higher-order spin correlations without conventional dipolar long-range order. Importantly, in the FM–AFM $J_1$–$J_2$ chain such bound states can emerge not only near saturation but also at intermediate magnetizations; in what follows we quantify their stability via the binding energy $E_{\rm b}(M,p)$.

\subsection{\label{subsec:BE}Definition of binding energy}

To quantify the stability of a bound state, we compute the magnon binding energy, defined as the energy gain of a composite excitation relative to separated magnons. For a $p$-magnon bound state at fixed magnetization $M$ (with conserved $S^z$), we use~\cite{J_Phys_Conf_Ser_400_032069}
\begin{align}
	E_{\rm b}(M,p)
	&= \big[E_0(S^z\!=\!M\!-\!1) - E_0(S^z\!=\!M)\big] \notag\\
	&\quad - \frac{1}{p}\big[E_0(S^z\!=\!M\!-\!p) - E_0(S^z\!=\!M)\big],
	\label{eq:be}
\end{align}
where $E_0(S^z)$ denotes the ground-state energy in the sector with $z$-component of total spin $S^z$. In what follows we identify $M$ with the total $S^z$ of a finite cluster, so that $M/M_{\rm s}$ denotes the normalized magnetization. When the largest positive value of $E_{\rm b}(M,n)$ in the thermodynamic limit is obtained for $n=p$, the $p$-magnon bound state ($p$-MBS) is the most stable one. If, on the other hand, $E_{\rm b}(M,n)=0$ for all $n$, no low-energy MBS exists. Since the most stable number of magnons is not known \textit{a priori}, $E_{\rm b}(M,p)$ is evaluated for several values of $p$. Note that Eq.~\eqref{eq:be} defines the binding energy \textit{per magnon}. All DMRG calculations conserve $S^z$ (fixed-magnetization sectors), and the ground-state energies $E_0(S^z)$ are obtained with high precision for use in Eq.~\eqref{eq:be}.

This criterion is fully consistent with the behavior of finite-size magnetization curves, where the formation of a $p$-MBS manifests itself as a magnetization step of width $\Delta S^z=p$~\cite{Heidrich-Meisner2006,Hikihara2008,Sudan2009}. In other words, $E_{\rm b}(M,p)>0$ corresponds to a regime in which $p$ magnons flip collectively, producing a single step in $M(h)$ instead of $p$ consecutive single-magnon steps $\Delta S^z=1$. Thus, the binding energy $E_{\rm b}(M,p)$ provides a quantitative measure of the same stability that can otherwise be qualitatively inferred from the discrete magnetization steps in finite clusters.

\subsection{\label{subsec:dmrg}Density-matrix renormalization group calculations}

We employ the DMRG method to evaluate (i) the zero-field pitch angle $\theta$ as a function of the frustration ratio $\alpha$, and (ii) the magnetization dependence of the magnon binding energy $E_{\rm b}(M,p)$.

\paragraph*{Pitch angle (OBC).}
In the spiral regime ($\alpha>1/4$), the pitch angle varies continuously with $\alpha$; enforcing periodic boundary conditions (PBC) would impose artificial commensurability constraints between the magnetic period and the lattice period. 
We therefore adopt open boundary conditions (OBC) and analyze bulk correlations.
Calculations are performed for chains up to $L=512$, and the results are extrapolated to the thermodynamic limit $L \to \infty$. 
Unless otherwise stated, up to $\chi=4000$ density-matrix eigenstates are retained, with typical discarded weights of $10^{-9}$–$10^{-10}$. 
The pitch angle $\theta$ is extracted from the oscillation wave vector determined from the peak position of the static spin structure factor, defined by
\begin{align}
S(q)=\frac{1}{L^2}\sum_{k,l=1}^L \langle \vec{S}_k \cdot \vec{S}_l \rangle \exp[iq(k-l)].
	\label{eq:sq}
\end{align}

\paragraph*{Binding energies (PBC).}
For the evaluation of excitation energies entering Eq.~\eqref{eq:be}, edge effects must be strictly controlled. This is particularly crucial in frustrated systems with competing length scales. We therefore use PBC for $E_{\rm b}(M,p)$. Near full saturation, when determining the $p$-magnon binding energy, system sizes are chosen as $L=pn$ ($n\!\in\!\mathbb{Z}^+$) to accommodate the $p$-magnon sector cleanly. For partial polarization we focus on $p=2$, $3$, and $4$. We then set $L=12n$ (the least common multiple of $2$, $3$, and $4$), which facilitates finite-size scaling. At fixed $\alpha$, the most energetically stable cluster size $p$ does not change upon lowering the magnetization within the MBS regime; accordingly, in Eq.~\eqref{eq:be} we evaluate sectors $M=M_{\rm s}-mp$ ($m \in \mathbb{Z}^+$) because of $\Delta S^z = p$ as mentioned above. PBC calculations are performed for systems up to $L=192$, and all results are extrapolated to $L \to \infty$. Up to $\chi=4000$ density-matrix eigenstates are retained, with largest discarded weights typically $\sim10^{-8}$. Where required, physical observables are extrapolated in $\chi$ prior to the thermodynamic extrapolation.

\section{\label{sec:results}Results}

In this section, we present numerical results for the formation and stability of MBSs in the spin-$\tfrac{1}{2}$ FM-AFM $J_1$-$J_2$ chain under magnetic fields. We first discuss the behavior of the zero-field pitch angle, which provides a useful reference for understanding the relationship between the spin configuration and the most stable bound-magnon size $p$. We then analyze the properties of bound states near full saturation and at intermediate magnetizations. Organization of this section is as follows: Sec.~\ref{subsec:pa} discusses the zero-field pitch angle, Sec.~\ref{subsec:saturation} analyzes bound states near saturation, and Sec.~\ref{subsec:intermediate} examines those at intermediate magnetizations.

\subsection{\label{subsec:pa}Pitch angle at $h=0$}

\begin{figure}[t]
	\includegraphics[width=0.7\linewidth]{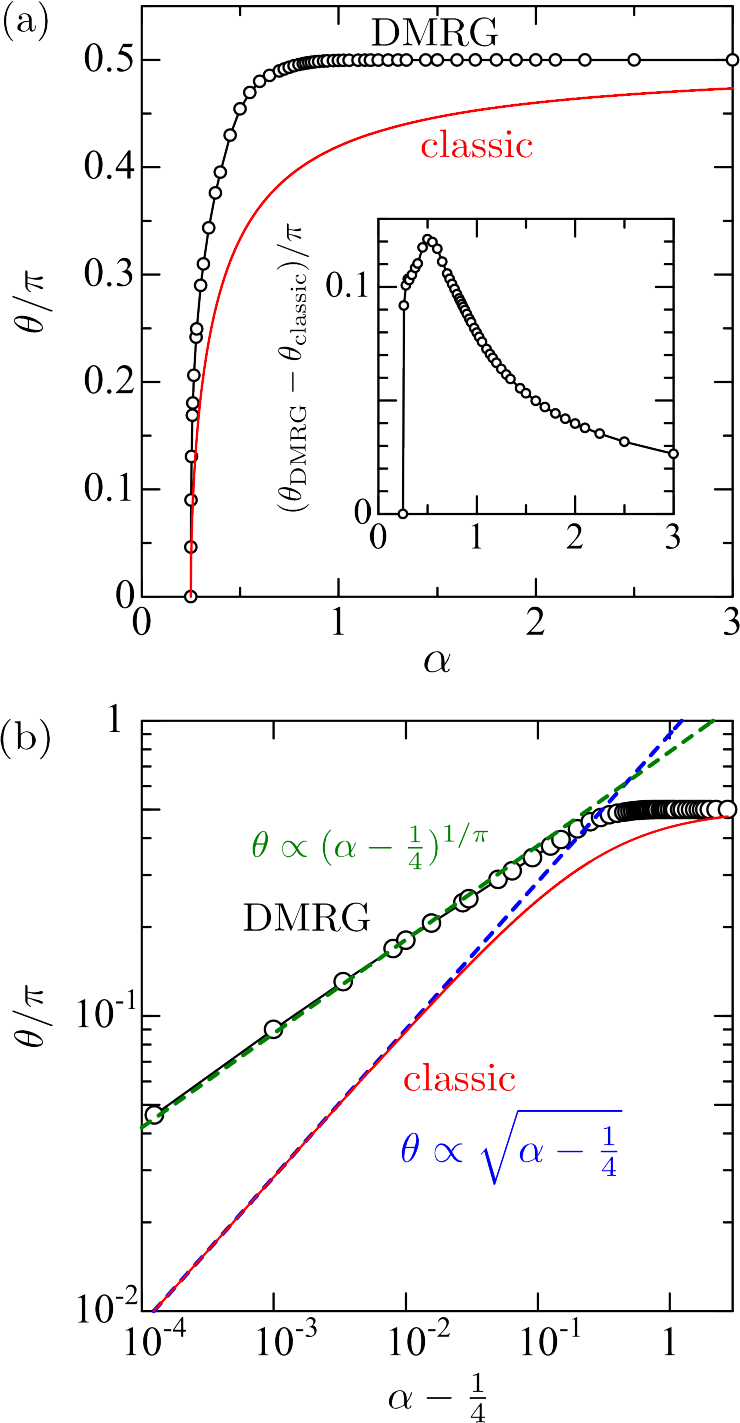}
	\caption{\label{fig:pa}
		(a) DMRG results for the pitch angle $\theta/\pi$ as a function of the frustration ratio $\alpha$. The classical result $\theta/\pi=\cos^{-1}(|J_1|/4J_2)/\pi$ is shown for comparison. Inset: Difference between the quantum and classical values of $\theta/\pi$.
		(b) Log–log plot of $\theta/\pi$ versus $\alpha-\tfrac14$, comparing the classical and DMRG scaling near the FM critical point.}
\end{figure}

It has been conjectured that the number of bound magnons in the high-field region is related to the pitch angle $\theta$ at zero field~\cite{Sudan2009}. As a preliminary step, we therefore examine the dependence of $\theta$ on $\alpha$. Figure~\ref{fig:pa}(a) shows the DMRG results for extrapolated values of $\theta/\pi$ to the thermodynamic limit, together with the classical prediction $\theta=\cos^{-1}[1/(4\alpha)]$. Details of the finite-size scaling analysis are provided in Appendix~\ref{app:scaling_pa}. In the FM phase ($\alpha<1/4$) the pitch angle remains $\theta=0$. Upon entering the spiral phase at $\alpha=1/4$, $\theta/\pi$ rapidly approaches $1/2$. The classical result exhibits a qualitatively similar trend, but its convergence to $\theta/\pi=1/2$ is significantly slower. The inset of Fig.~\ref{fig:pa}(a) displays the deviation $\Delta\theta/\pi$ between the DMRG and classical values as a function of $\alpha$, which peaks around $\alpha\!\sim\!0.5$, indicating that quantum fluctuations are most pronounced in this regime~\cite{Comment2011}. 

Interestingly, $\Delta\theta/\pi$ shows an almost discontinuous rise near the transition point. Although both the DMRG and classical pitch angles are continuous functions of $\alpha$, their onsets differ drastically. To clarify this difference, Fig.~\ref{fig:pa}(b) presents log–log plots of $\theta/\pi$ versus $\alpha-1/4$. Both results display singular behavior near the transition. Expanding the classical expression around $\alpha=1/4$ yields
\[
\theta \simeq 2\sqrt{2}\,\sqrt{\alpha - \tfrac{1}{4}} + \mathcal{O}\!\left[(\alpha-1/4)^{3/2}\right],
\]
indicating $\theta\propto \delta^{1/2}$ with $\delta\equiv \alpha-1/4$. In contrast, the DMRG data are well fitted by
\[
\theta \simeq \frac{\pi^2}{4}\,(\alpha- \tfrac{1}{4})^{1/\pi},
\]
showing a much slower, nonclassical critical behavior.

Near the Lifshitz point $\alpha_c=1/4$, let $\delta=\alpha-\alpha_c\to 0^+$. Classically, the long-wavelength energy expansion,
\[
\mathcal{E}(q)=\tfrac12\rho(\delta)\,q^2+\kappa q^4+\cdots,
\]
with $\rho(\delta)\propto\delta$, leads to $\theta\sim q\propto \delta^{1/2}$. For the quantum spin-$\tfrac12$ chain, however, the $z=4$ \emph{quantum Lifshitz} criticality and Berry-phase effects render the effective stiffness nonanalytic, $\rho_{\mathrm{eff}}(\delta)\propto\delta^{2/\pi}$, resulting in $\theta\sim q\propto\delta^{1/\pi}$, in excellent agreement with the DMRG data~\cite{Sirker2011,Balents2016}. Equivalently, the incommensurability measures the density of chiral solitons forming a Tomonaga–Luttinger liquid; with an effective Luttinger parameter $K\simeq 2/\pi$ near $\alpha_c$, one expects $q\propto\delta^{K/2}=\delta^{1/\pi}$. This explains the nearly discontinuous rise of $\theta/\pi$ while keeping $\theta$ itself continuous at $\alpha_c$.

Tabulated values of $\theta/\pi$ for various $\alpha$ are provided in Appendix~\ref{app:data_pa} for convenience. Because the pitch angle is experimentally accessible (e.g., via inelastic neutron scattering or ESR), it provides a direct handle to determine the \emph{effective} frustration ratio $\alpha$ in materials where $J_1$ and $J_2$ dominate the exchange network. In this sense, the theoretical relation between $\theta$ and $\alpha$ established here offers a practical framework for extracting an effective $\alpha$ that implicitly incorporates renormalization effects from smaller interactions. We note that, particularly in the FM-AFM $J_1$-$J_2$ chain, even very tiny interchain couplings can have a substantial impact on the effective $\alpha$~\cite{Saturation2011,EPL2012,Willenberg2016}.

\subsection{\label{subsec:saturation}Near saturation}

\begin{figure}[t]
	\includegraphics[width=0.7\linewidth]{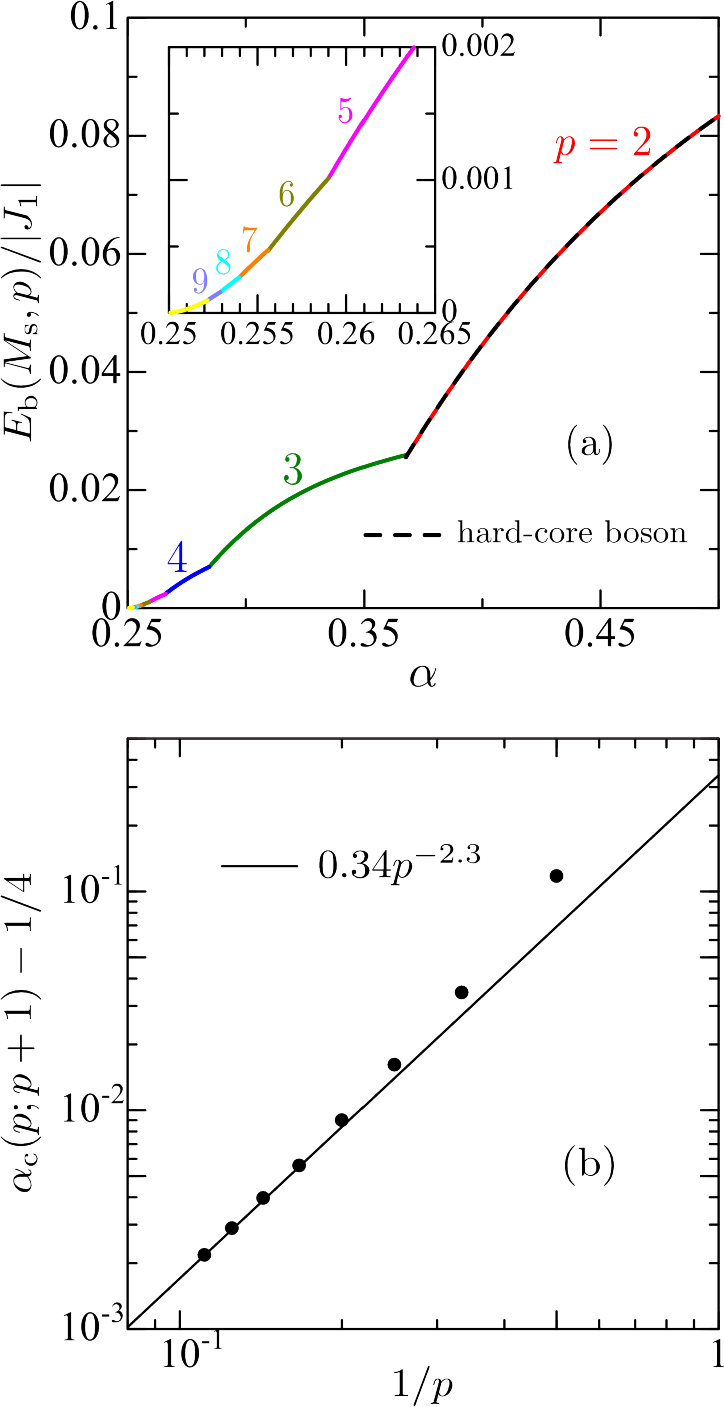}
	\caption{\label{fig:Eb_Ms}
		(a) DMRG results for the binding energy $E_{\rm b}(M,p)$ near full saturation ($M/M_{\rm s}=1$) as a function of $\alpha$. The most stable bound-magnon size $p$ is indicated for each region.
		(b) Critical coupling $\alpha_{\rm c}(p;p+1)-1/4$ between the $p$- and $(p\!+\!1)$-MBS phases plotted as a function of $1/p$. The solid line shows a fit to $0.34\,p^{-2.3}$.}
\end{figure}

In the fully polarized state induced by an external magnetic field, MBSs appear in the frustrated regime $\alpha>1/4$. The number of bound magnons $p$ depends sensitively on $\alpha$ and can be determined by evaluating the binding energy $E_{\rm b}(M,p)$ defined in Eq.~\eqref{eq:be}.

Figure~\ref{fig:Eb_Ms}(a) shows the DMRG results for $E_{\rm b}(M,p)$ near full saturation ($M/M_{\rm s}=1$) as a function of $\alpha$, with the corresponding bound-magnon number $p$ indicated. In the region close to the FM critical point ($0.25<\alpha<0.5$), the binding energy decreases as $\alpha$ approaches $1/4$. For the simplest case $p=2$, the binding energy can be obtained analytically using the hard-core boson approach as~\cite{Kuzian2007}
\begin{align}
	E_{\rm b}(M_{\rm s},2)=\frac{3\alpha-1}{8\alpha(\alpha+1)},
	\label{eq:Eb2}
\end{align}
which reaches its maximum at $\alpha=1$. The DMRG results are in excellent agreement with this analytic expression.

Phase transitions between different $p$-MBS states are identified by direct crossings of the binding energies $E_{\rm b}(M_{\rm s},p)$ for successive $p$ values, which appear as kinks in the energy curves. As $p$ increases, the stability window of each $p$-MBS phase becomes progressively narrower. To quantify this trend, the critical couplings $\alpha_{\rm c}(p;p\!+\!1)$ separating the $p$- and $(p\!+\!1)$-MBS phases are plotted against $1/p$ in Fig.~\ref{fig:Eb_Ms}(b). For large $p$, the data follow the empirical scaling relation $\alpha_{\rm c}(p;p\!+\!1)-1/4 \simeq 0.34\,p^{-2.3}$, indicating a hierarchical sequence of bound states systematically approaching the FM critical point.

\begin{figure}[t]
	\includegraphics[width=0.8\linewidth]{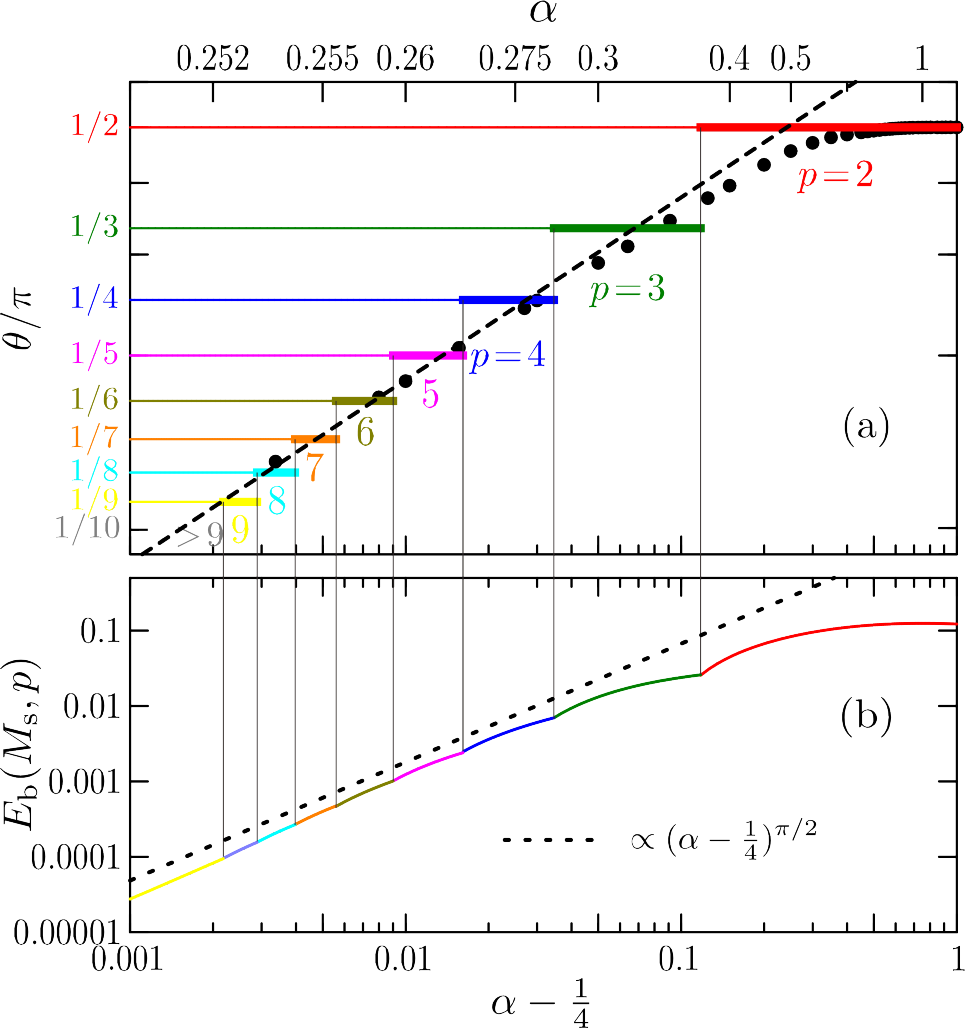}
	\caption{\label{fig:Eb_pa}
		(a) Pitch angle $\theta/\pi$ (black circles) at zero field as a function of $\alpha-1/4$, together with the corresponding most energetically stable (MES) bound-magnon number $p$ at saturation. The dashed line denotes the quantum critical scaling $\theta/\pi=(\pi/4)(\alpha-1/4)^{1/\pi}$.
		(b) Magnon binding energy $E_{\rm b}(M_{\rm s},p)$ corresponding to $p$ in the MES at saturation. The dotted line represents an \emph{empirical} power-law fit $E_{\rm b}(M_{\rm s},p) \propto (\alpha-1/4)^{\pi/2}$.}
\end{figure}

It has been suggested that the conjectured relationship between $\theta$ and $p$~\cite{Sudan2009},
\begin{align}
	\frac{1}{p} > \frac{\theta}{\pi} > \frac{1}{p+1},
	\label{eq:theta_pa}
\end{align}
holds over a wide range of $\alpha$. Figure~\ref{fig:Eb_pa}(a) shows the zero-field pitch angle $\theta/\pi$ and the number of bound magnons $p$ near full saturation ($M/M_{\rm s}=1$) as a function of $\alpha-1/4$. The conjecture of Eq.~\eqref{eq:theta_pa} is well satisfied at least up to $p\!\lesssim\!9$. The number $p$ can be intuitively interpreted as the number of spins contained in one lobe (half period) of the incommensurate spin modulation. The corresponding binding energies in the thermodynamic limit are shown in Fig.~\ref{fig:Eb_pa}(b). Approaching the FM instability, the bound states become larger (larger $p$) while their binding energies are strongly suppressed, indicating that extended bound states are increasingly fragile. Close to the critical point, $E_{\rm b}(M_{\rm s},p)$ follows an empirical power-law vanishing $E_{\rm b}(M_{\rm s},p)\propto (\alpha-1/4)^{\pi/2}$.

\subsection{\label{subsec:intermediate}Intermediate magnetization}

\begin{figure}[t]
	\includegraphics[width=0.9\linewidth]{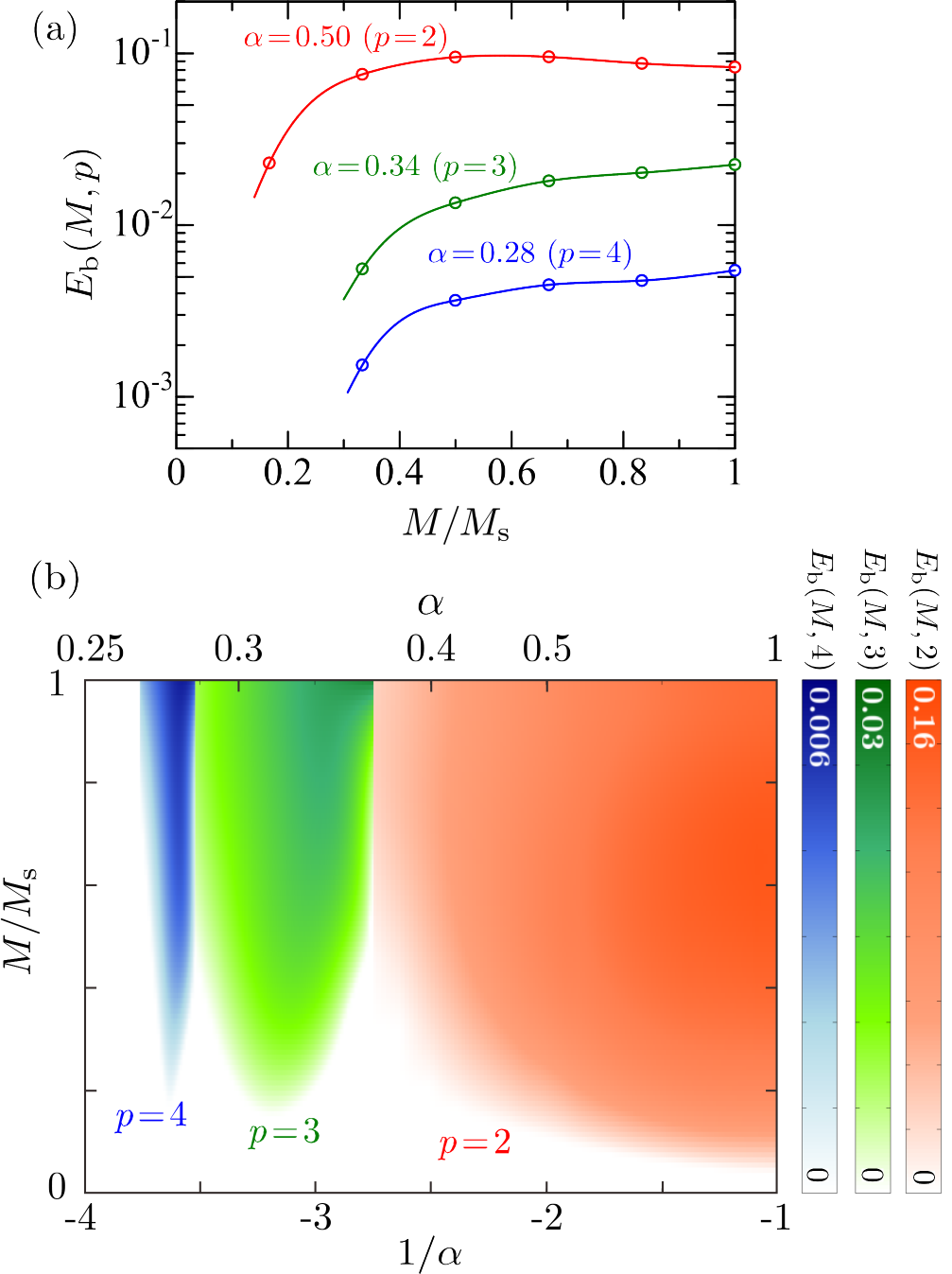}
	\caption{\label{fig:Eb_Mdep}
		(a) Representative cross sections of the binding energy $E_{\rm b}(M,p)$ as a function of $M/M_{\rm s}$ for selected frustration ratios $\alpha=0.50$, $0.34$, and $0.28$, which correspond to the nematic ($p=2$), triatic ($p=3$), and quartic ($p=4$) states, respectively.
		(b) Color map of $E_{\rm b}(M,p)$ obtained by DMRG as a function of $1/\alpha$ ($\alpha$) and $M/M_{\rm s}$.}
\end{figure}

We now examine how the binding energy evolves below full saturation ($0<M/M_{\rm s}<1$). Figure~\ref{fig:Eb_Mdep}(a) shows $E_{\rm b}(M,p)$ at $\alpha=0.50$, $0.34$, and $0.28$, which correspond to the nematic ($p=2$), triatic ($p=3$), and quartic ($p=4$) states at full saturation, respectively. For $\alpha=0.50$, the binding energy remains nearly unchanged upon reducing the magnetization from $M/M_{\rm s}=1$ down to $\sim 0.2$, followed by a rapid collapse to zero at lower fields. Notably, in this case $E_{\rm b}(M,2)$ attains its maximum around $M/M_{\rm s}\!\approx\!0.6$, rather than exactly at saturation. This indicates that partial depolarization enhances the kinetic motion of bound magnons, allowing them to gain propagation energy along or against the field. For $\alpha=0.34$ and $0.28$, $E_{\rm b}(M,p)$ decreases gradually with decreasing magnetization and then drops sharply to zero at a characteristic field, implying that bound magnons survive as well-defined quasiparticles down to relatively low magnetizations in our data. These trends suggest a crossover from a dilute magnon gas near saturation to a more densely interacting bound-magnon liquid at lower fields. These trends are consistent with the behavior of the lowest excitation gap in the dynamical spin structure factor, which represents the energy required to break a bound magnon pair or  cluster~\cite{Onishi2015}.

The evolution of the most energetically stable cluster size $p$ with decreasing $M/M_{\rm s}$ provides additional insight into the hierarchy of multipolar correlations. Figure~\ref{fig:Eb_Mdep}(b) summarizes $E_{\rm b}(M,p)$ over the $(\alpha,\,M/M_{\rm s})$ plane. Smaller-$p$ states persist to comparatively lower magnetizations, whereas the stability windows of larger-$p$ states shrink rapidly and their boundaries bend noticeably with $M$. It is also noteworthy that, for a fixed $\alpha$, the value of $p$ remains unchanged as $M$ varies, indicating that the character of the bound state is essentially determined by the frustration ratio. Overall, the regions where $E_{\rm b}(M,p)$ remains finite for $p=2$, $3$, and $4$ are in good accord with the multipolar (SDW$_p$) domains of the field-dependent phase diagram reported in Refs.~\cite{Hikihara2008,Sudan2009}.

\section{\label{subsec:conclusion}Conclusion}

We have systematically investigated the stability of MBSs in the spin-$\tfrac12$ FM-AFM $J_1$-$J_2$ chain under magnetic fields by computing the binding energy $E_{\rm b}(M,p)$ using the DMRG method. In the near-saturation regime, we identified a clear hierarchy of $p$-magnon phases that follows an empirical scaling of the phase boundaries, $\alpha_{\rm c}(p;p{+}1)-\tfrac{1}{4}\!\simeq\!0.34\,p^{-2.3}$ for large $p$. The proposed correspondence between the most stable cluster size $p$ and the zero-field pitch angle $\theta$ was confirmed over a broad frustration window, validating the inequality $1/p>\theta/\pi>1/(p{+}1)$ up to $p\!\lesssim\!9$. Approaching the FM critical point $\alpha=1/4$, the binding energy per magnon is strongly reduced and, for certain $\alpha$, reaches its maximum slightly below full saturation, indicating that partial depolarization enhances the kinetic motion of bound magnons. 

At intermediate magnetizations, bound clusters persist as well-defined quasiparticles down to relatively low $M/M_{\rm s}$, implying a crossover from a dilute magnon gas near saturation to a more densely interacting bound-magnon liquid at lower fields. The comprehensive $E_{\rm b}(M,p)$ maps, together with the tabulated $\theta(\alpha)$ values and finite-size scaling analyses in the Appendices, provide concrete reference data for inelastic neutron scattering, ESR/NMR relaxation, and high-field magnetization experiments on Q1D frustrated magnets.

Future extensions may include the effects of exchange anisotropy, Dzyaloshinskii–Moriya interactions, interchain coupling, and finite temperature, as well as systems with internally induced magnetization. Such investigations will further clarify the robustness of MBSs beyond the idealized isotropic chain and guide the search for their unambiguous realization in real materials.

{\it Acknowledgements.---}
We thank Ulrike Nitzsche for technical support. This project is funded by the German Research Foundation (DFG) via the projects A05 of the Collaborative Research Center SFB 1143 (project-id 247310070). This work is dedicated to the memory of Johannes Richter, with whom we had the privilege of many fruitful collaborations.

\vspace*{5.0mm}

The data that support the findings of this study are available from the author upon reasonable request.

\appendix

\section*{Appendix}

\begin{table}[t]
	\caption{\label{tab:pitch}
		DMRG results for the zero-field pitch angle $\theta/\pi$ 
		as a function of the frustration ratio $\alpha = J_2/|J_1|$. 
		In the limit $\alpha \to \infty$, the pitch angle approaches $\theta/\pi \to 1/2$.}
	\begin{ruledtabular}
		\begin{tabular}{cccc}
			$\alpha$ & $\theta/\pi$ (DMRG) & $\theta/\pi$ (classical) & deviations \\ \hline
			0.26 & 0.18036 & 0.08857 & 0.09179 \\
			0.28 & 0.24935 & 0.14870 & 0.10065 \\
			0.30 & 0.28992 & 0.18643 & 0.10349 \\
			0.40 & 0.39536 & 0.28510 & 0.11026 \\
			0.50 & 0.45430 & 0.33333 & 0.12097 \\
			0.75 & 0.49307 & 0.39183 & 0.10124 \\
			1.00 & 0.49939 & 0.41956 & 0.07983 \\
			2.00 & 0.49990 & 0.46011 & 0.03979 \\
			3.00 & 0.49992 & 0.47344 & 0.02648 \\
		\end{tabular}
	\end{ruledtabular}
\end{table}

\section{Tabulated DMRG data for the pitch angle}\label{app:data_pa}

Table~\ref{tab:pitch} lists the zero-field pitch angle $\theta$ (in units of $\pi$) obtained by DMRG for various values of the frustration ratio $\alpha = J_2/|J_1|$. The values correspond to the thermodynamic-limit estimates shown in Fig.~\ref{fig:pa}. Finite-size scaling analysis of the pitch angle is performed in Appendix~\ref{app:scaling_pa}. For comparison, the classical values are also included. The deviation between the DMRG and classical values is largest around $\alpha\!\sim\!0.5$, indicating that quantum fluctuations are most pronounced in this regime~\cite{Comment2011}. These data may serve as a reference for estimating effective $\alpha$ values from experimentally observed pitch angles. It should be noted that the effective $\alpha$ thus inferred implicitly includes the renormalization effects of all smaller interactions in real materials. In Q1D systems, such weak interactions can have a significant impact on the determination of the effective $\alpha$. For instance, even when the bare $\alpha$ lies on the spiral side ($\alpha>1/4$), the renormalization due to small interchain couplings or other perturbations may shift the effective $\alpha$ to the ferromagnetic side ($\alpha<1/4$)~\cite{Saturation2011,Kuzian2018}.

\section{Finite-size scaling analysis for the pitch angle}\label{app:scaling_pa}

\begin{figure}[t]
	\includegraphics[width=0.7\linewidth]{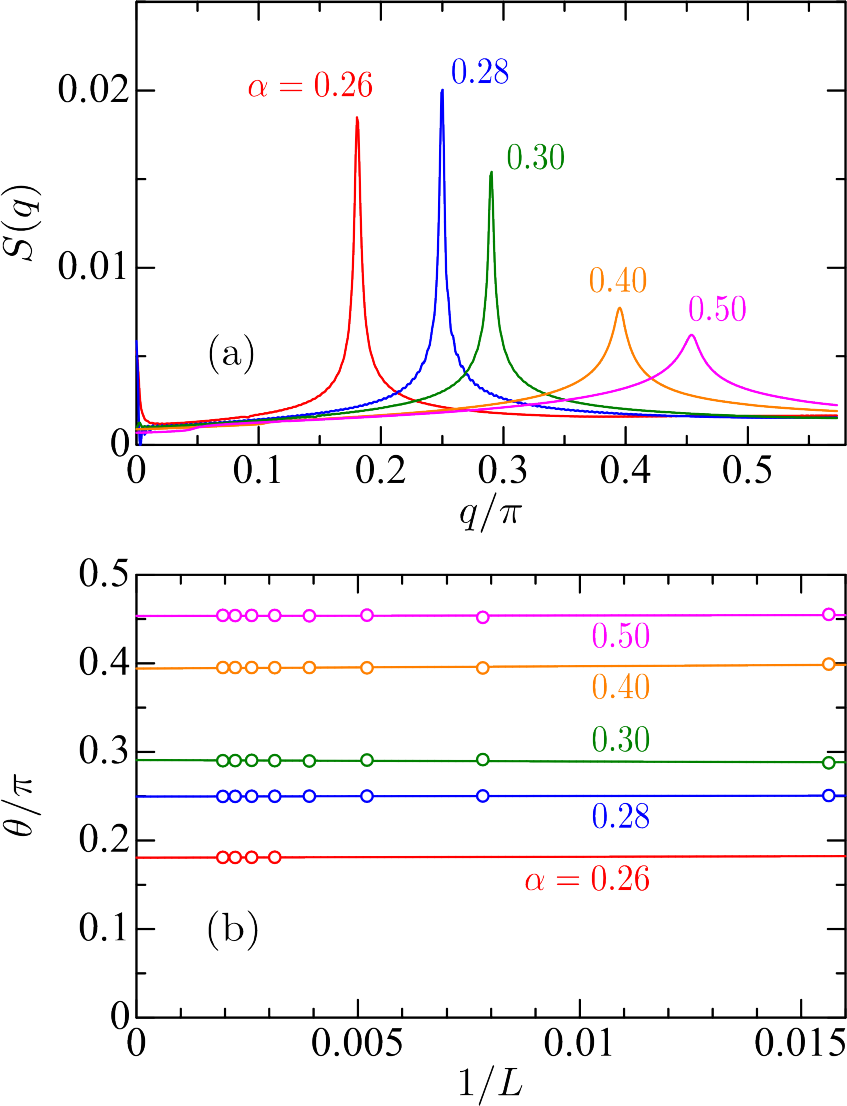}
	\caption{\label{fig:scaling_pa_App}
		(a) Static spin structure factor $S(q)$ obtained for open chains with $L=512$ at several values of $\alpha>1/4$. 
		(b) Examples of finite-size scaling of $\theta$ for selected $\alpha$ values. A polynomial function is used for the fitting.
	}
\end{figure}

Here we briefly describe the procedure used to extract the pitch angle and show several examples of its extrapolation to the thermodynamic limit. In this study, the static spin structure factor $S(q)$ defined in Eq.~\eqref{eq:sq} is calculated, and the pitch angle $\theta$ is determined from the position of its dominant peak. As discussed in the main text, we focus on the incommensurate spiral regime ($\alpha>1/4$) of the FM-AFM $J_1$-$J_2$ chain and therefore employ OBC to avoid artificial commensurability constraints between the magnetic and lattice periods.

Figure~\ref{fig:scaling_pa_App}(a) displays representative $S(q)$ data obtained for open chains with $L=512$ at several $\alpha$ values. For each $\alpha$, a single well-defined peak is observed. Since the spiral state lacks true long-range order, the peak exhibits a finite width; however, it becomes progressively sharper as $\alpha$ approaches the FM critical point ($\alpha=1/4$), reflecting the suppression of quantum fluctuations. Figure~\ref{fig:scaling_pa_App}(b) presents examples of the finite-size scaling of $\theta$ for selected $\alpha$ values. Because the size dependence is relatively weak, a smooth extrapolation to the thermodynamic limit can be achieved using polynomial fits. The extrapolated values of $\theta$ are summarized in Fig.~\ref{fig:Eb_pa} and Table~\ref{tab:pitch}.

\section{Finite-size scaling analysis for the binding energy}\label{app:scaling_be}

\begin{figure}[b]
	\includegraphics[width=1.0\linewidth]{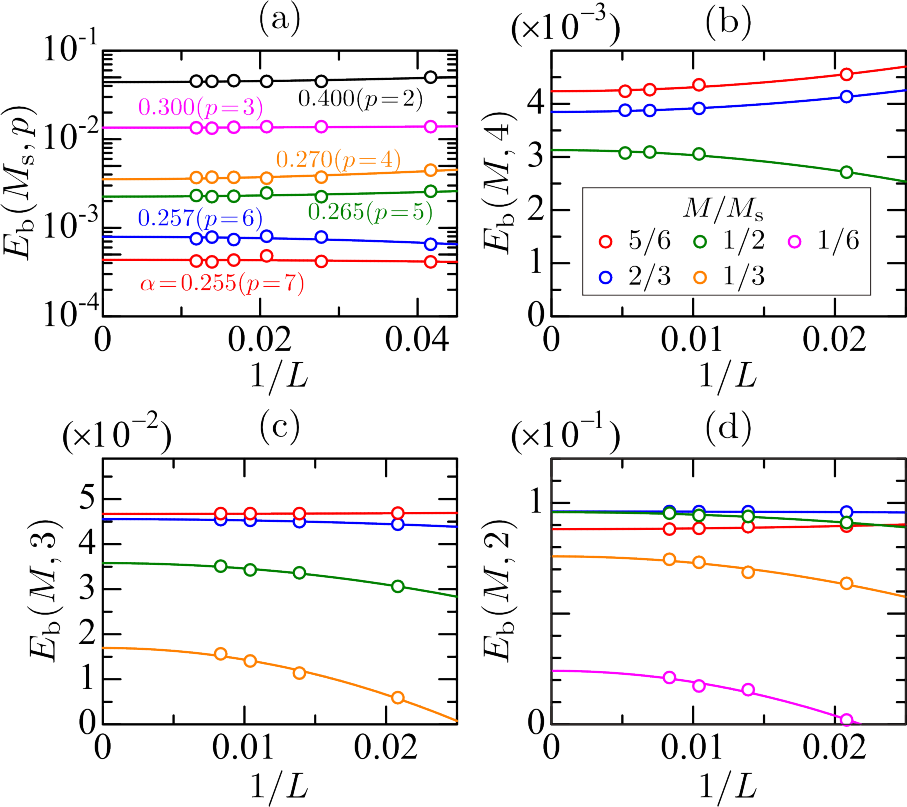}
	\caption{\label{fig:scaling_Eb_App}
		(a) Finite-size scaling of the binding energy $E_{\rm b}(M_{\rm s},p)$ at full polarization for several frustration ratios. Periodic chains up to $L=80$ sites were used. 
		(b--d) Examples of finite-size scaling of $E_{\rm b}(M,p)$ away from full saturation for $\alpha=0.276$ ($p=4$), $0.30$ ($p=3$), and $0.40$ ($p=2$). Periodic chains up to $L=144$ sites were used. Solid lines represent fits to $E_{\rm b}(M,p)=\beta+\gamma/L^2$, where $\beta$ and $\gamma$ are fitting parameters.}
\end{figure}

Here we present several examples of the finite-size scaling analysis for the magnon binding energy. Figure~\ref{fig:scaling_Eb_App}(a) shows the finite-size scaling of $E_{\rm b}(M_{\rm s},p)$ at full polarization for various values of $\alpha$, with the corresponding $p$ values indicated in the figure. As discussed in the main text, PBC were employed to eliminate unexpected edge effects in the estimation of excitation energies. Periodic chains with lengths up to $L=80$ were used in these calculations. Overall, the size dependence is weak, and a smooth extrapolation to the thermodynamic limit is straightforward.

Figures~\ref{fig:scaling_Eb_App}(b--d) display the scaling behavior of $E_{\rm b}(M,p)$ away from full saturation for $\alpha=0.276$ ($p=4$), $0.30$ ($p=3$), and $0.40$ ($p=2$). Periodic chains with lengths up to $L=144$ were used in these calculations. As the magnetization decreases, the spin \textit{mobility} increases, effectively broadening the magnon band. This results in somewhat stronger finite-size effects, but in all cases the data are well fitted by $E_{\rm b}(M,p)=\beta+\gamma/L^2$, where $\beta$ and $\gamma$ are fitting parameters.

\bibliography{ZNA}

\end{document}